\documentclass{aa}  

\usepackage{natbib}
\usepackage{graphicx}
\usepackage{txfonts}
\newcommand{\be}{\begin{equation}}
\newcommand{\ee}{\end{equation}}
\newcommand{\ba}{\begin{eqnarray}}
\newcommand{\ea}{\end{eqnarray}}

\begin{document}


\title{Detection asymmetry in solar energetic particle events}

\titlerunning{Corotation in SEP events}

   \author{S.~Dalla\inst{1}
          \and
          A.~Hutchinson\inst{1,2,3}
                   \and
           R.A.~Hyndman\inst{1} 
         \and
           K.~Kihara\inst{4,5} 
         \and
         N.V.~Nitta\inst{6}
          \and
          L.~Rodr{\'\i}guez-Garc{\'\i}a,\inst{7,8}
          \and
           T.~Laitinen\inst{1} 
          \and
           C.O.G.~Waterfall\inst{1,2,9} 
           \and 
                     D.S.~Brown\inst{1}
           }

   \institute{Jeremiah Horrocks Institute, University of Central Lancashire,
    Preston, PR1 2HE, UK \\
    \email{sdalla@uclan.ac.uk}
    \and
    Heliophysics Division, NASA Goddard Space Flight Center, Greenbelt, MD 20771, USA
    \and
    Department of Physics, University of Maryland, Baltimore County, 1000 Hilltop Circle, Baltimore, MD 21250, USA
    \and
    Secure System Platform Research Laboratories, NEC Corporation, Kanagawa 211-8666, Japan
    \and 
    Astronomical Observatory, Kyoto University, Sakyo, Kyoto 606-8502, Japan
    \and
        Lockheed Martin Advanced Technology Center, 3251 Hanover Street, Palo Alto, CA 94304, USA    
        \and
        European Space Agency (ESA), European Space Astronomy Centre (ESAC), Camino Bajo del Castillo s/n, 28692 Villanueva de la Cañada, Madrid, Spain
        \and
        Universidad de Alcalá, Space Research Group (SRG-UAH), Plaza de San Diego s/n, 28801 Alcalá de Henares, Madrid, Spain
        \and
        University Corporation for Atmospheric Research, 3090 Center Green Dr., Boulder, CO 80301, USA
   }

   \date{Received November 2024; }

 
  \abstract
   {Solar energetic particles (SEPs) are detected in interplanetary space in association with solar flares and coronal mass ejections (CMEs). The magnetic connection between the observing spacecraft and the solar active region (AR) source of the event is a key parameter in determining whether SEPs are observed and the particle event's properties.}
   {We investigate whether an east-west asymmetry in the detection of SEP events is present in observations and discuss its possible link to  corotation of magnetic flux tubes with the Sun.}
   {We used a published dataset of 239 CMEs recorded between 2006 and 2017 and having source regions both on the Sun's front and far sides as seen from Earth.  We produced distributions of occurrence of in-situ SEP intensity enhancements associated with the CME events, versus $\Delta \phi$, the longitudinal separation between source active region and spacecraft magnetic footpoint based on the nominal Parker spiral. We focused on protons of energy $>$10 MeV measured by STEREO A, STEREO B and GOES at 1 au. We also considered  occurrences of 71-112 keV electron events  detected by MESSENGER between 0.31 and 0.47 au. }
   {We find an east-west asymmetry with respect to the best magnetic connection ($\Delta \phi$=0)  in the detection of $>$10 MeV proton events and of 71-112 keV  electron events. For protons, observers for which the source AR is on the east side of the spacecraft footpoint  and not well connected ($-$180$^{\circ}$$<$$\Delta \phi$$<$$-$40$^{\circ}$) are 93\% more likely to detect an SEP event compared to observers with $+$40$^{\circ}$$<$$\Delta \phi$$<$$+$180$^{\circ}$.  The asymmetry may be a signature of corotation of magnetic flux tubes with the Sun, since for events with $\Delta \phi$$<$0 corotation sweeps particle-filled flux tubes towards the observing spacecraft, while for $\Delta \phi$$>$0 it takes them away from it. Alternatively it may be related to asymmetric acceleration or propagation effects.}
   {}


   \maketitle
%
%

\section{Introduction}\label{sect_intro}

Solar energetic particles (SEPs), accelerated as a result of energy release events at the Sun, are detected by spacecraft instruments in interplanetary space in close temporal coincidence with flares and coronal mass ejections (CMEs). Time-intensity profiles of electrons, protons and heavy ions have been characterised extensively over several decades of particle observations (e.g. \citealt{vanHol1975, Can1988a, Ric2014, Pap2016, Coh2017, Rod2023a}). 
During so-called gradual SEP events, intensities measured at 1 au often remain elevated above background for several days (e.g. \citealt{Des2016,Kle2017,Coh2021}).

A striking feature of SEP observations is the so-called east-west (E-W) effect in the particle intensity profiles: for a near-Earth spacecraft events with source active region (AR) in the west of the Sun tend to have a fast rise to peak intensity and decay, while those with source AR in the east typically have a slow rise and longer duration \citep{Can1988a}.  \citet{vanHol1975} were the first to analyse the dependence of SEP profile characteristics, such as the rise time and spectral index, on the longitude of the source AR, showing the presence of E-W asymmetries. Asymmetries have been confirmed by  a number of more recent studies (e.g.~\citealt{Lar2013,Ric2014}).
In a study of electron SEP events, \citet{Rod2023b} noted an asymmetry to the east in the range of $\Delta \phi$ values for which the highest peak intensities are observed. Here $\Delta \phi$, sometimes termed connection angle, gives the difference in longitude between the source AR and the observer's magnetic footpoint at the Sun,  so that $\Delta \phi$$<$0 indicates an AR east of the magnetic footpoint  (note that some studies, e.g.~\citealt{Ric2014}, use a definition with opposite sign).
\citet{Can1988a} proposed that the qualitative dependence of SEP intensity profiles on the location of the source AR is the result of different geometries of magnetic connection of the observer to the CME-driven shock.
In this interpretation  particle are accelerated at the shock with a spatially varying acceleration efficiency along its front so that the particle intensity at injection depends on which portion of the shock front is connected to the observer
(as discussed also by \citealt{Tyl2005}). \citet{Din2022} interpreted the east-west asymmetry in SEP fluence as due to the combined effect of the shock acceleration history and the geometry of the interplanetary magnetic field.

In addition to E-W asymmetries in the parameters of SEP intensity profiles, there have been indications in the literature of 
a longitude asymmetry in the detection of SEP events.
In a study using data from the Helios 1 and Helios 2 spacecraft gathered between 1974 and 1985, \citet{Kal1992b} noted that for the 77 SEP events they analysed, approximately 2/3 of the events had $\Delta \phi$$<$0, though they commented that this was unlikely to result from a real physical mechanism and was likely due to spacecraft orbits. During the time range considered
in their study, the Helios spacecraft were magnetically connected to the far side of the Sun for part of the time, while observations of flares were available for the front side only. 
\citet{Dal2003} studied a subset of the same events in two proton and one electron channels and plotted the event duration versus  $\Delta \phi$, noting that the results displayed an E-W asymmetry in duration. They also commented that in this dataset, events associated with  large negative $\Delta \phi$ were much more likely than those with large positive $\Delta \phi$ and showed that this could not be ascribed to spacecraft trajectories, terming the effect \lq detection asymmetry\rq. They suggested that corotation may help explaining the observations. Using a list of 78 solar proton events affecting the Earth environment collected by the NOAA Space Weather Prediction Centre during 1996 to 2011,  \citet{He&Wan_2017} noted an excess of events with negative  $\Delta \phi$, for   $| \Delta \phi |$$\lesssim$40$^{\circ}$.

As the solar wind propagates radially outwards and the footpoints of magnetic field lines remain anchored at the photosphere, the Parker spiral structure of the interplanetary magnetic field (IMF) is generated.
Magnetic flux tubes in the heliosphere appear to corotate with the Sun, an effect that is evident in movies of simulations of the solar wind structure from models such as ENLIL \citep{Ods1999}, EUHFORIA \citep{Pom2018} and Huxt \citep{Owe2020}. Observations in the heliosphere demonstrated the presence of features recurring at the solar rotation period (e.g.~\citealt{Heb1999, For2001}).
Corotation is very important in shaping measured properties of the solar wind at 1 au, as demonstrated by the success of empirical solar wind forecast models based on it \citep{Owe2013}.

From the point of view of an inertial (non-corotating) frame, once SEPs have been injected into the heliosphere, corotation sweeps particle-filled magnetic flux tubes away from/towards an observer. In many cases the same is true in the spacecraft frame, since the velocity of a 1-au spacecraft is small compared to the corotation and solar wind velocities (exceptions may be spacecraft located very close to the Sun,  for example Parker Solar Probe and Solar Orbiter for part of their orbits). 
In early SEP studies corotation was thought to be important to explain SEP events \citep{Bur1967}. For so-called impulsive SEP events, thought to be produced by solar flares, simulations have shown that corotation affects the modelled intensity profiles
\citep{Gia2012, Dro2010}.
However for SEP events resulting from acceleration at CME-driven shocks, 1D focussed transport models which included corotation in
an approximate way concluded it has negligible effects \citep{Kal1997, Lar1998}. As a result, the influence of corotation on gradual events is regarded as minimal and generally it is neglected.
Corotation is not routinely included in SEP focussed transport models nor forecasting tools \citep{Whi2023}.

Recent models based on 3D test particle simulations reached a very different conclusion: \citet{Mar2015} described the formation of corotating SEP streams, particle-filled magnetic flux tubes that corotate with the Sun. Modelling SEPs injected instantaneously at the Sun, they noted that in test particle simulations the E-W effect in SEP intensity profiles develops naturally as a result of corotation, as was also pointed out by \citet{Dal2017a}. Simulations including time-extended acceleration from a wide shock-like source reached the same conclusion \citep{Hut2023a}.
Thus, according to test particle simulations, corotation effects are important in shaping SEP intensity profiles for both impulsive and gradual events.
Using a simple 1D diffusion model and an impulsive and wide injection at the Sun, \citet{Lai2018} demonstrated the qualitative differences in the intensity profiles of 10 MeV protons from a model that included corotation and one that did not, for a scattering mean free path $\lambda$=0.03 au.
It remains to be established whether  any signatures of corotation are visible in SEP observations and whether it plays a role in  E-W asymmetries.

Regarding a possible SEP detection asymmetry, so far it has been difficult to characterise it conclusively.
One reason for this is that in most studies in the past both flare and SEP observations were affected by Earth bias:  for the majority of events, only source regions on the front side of the Sun were identified routinely via the associated flare and only spacecraft near the Earth measured SEPs.
Due to the winding of the IMF, for example assuming a solar wind speed of 450 km s$^{-1}$, the footpoint of a near-Earth spacecraft is located at longitude $\phi_{\mathit{ftpt}}$=55$^{\circ}$ with respect to the Earth-Sun line. Thus a source AR at the west limb gives  $\Delta \phi$=35$^{\circ}$ and larger positive  $\Delta \phi$ values are not accessible if only front side source regions are used, for a near-Earth spacecraft.
Thus analysis of the entire  [$-$180$^{\circ}$,$+$180$^{\circ}$] range of  $\Delta \phi$ values was not possible. The situation changed  thanks to the Solar TErrestrial RElations Observatory (STEREO) mission, consisting of two spacecraft orbiting the Sun at about 1 au, one moving ahead of the Earth and one behind it \citep{Kai2008}. STEREO data allowed identification of source active regions on the far side of the Sun via the Extreme UltraViolet Imager (EUVI) instrument  \citep{Wue2004}, as well as SEP detection when the spacecraft were magnetically connected to regions on the far side of the Sun.

In this paper we address the question of whether indications from previous studies of an E-W asymmetry in SEP event detection can be confirmed, by using a large statistical sample of CMEs with accurate information on their source regions. This is a  previously published dataset of 239 front-side and far-side CME events that took place during the STEREO era \citep{Kih2020}. Within the dataset, the SEP effects, if any, of the events were identified by analysing $>$10 MeV proton data from STEREO A, STEREO B and the Geostationary Operational Environmental Satellites (GOES). We derive $\Delta \phi$ distributions of SEP detections from this dataset to show that an E-W detection asymmetry with respect to $\Delta \phi$=0 (nominal best magnetic connection) is present. We also present distributions of detections of 71-112 keV electron events at the MErcury Surface Space ENvironment GEochemistry and Ranging (MESSENGER) spacecraft \citep{Sol2007}, located at radial distances between 0.31 and 0.47 au, making use of the dataset of 61 events from  \citet{Rod2023a}.
We discuss whether corotation may play a role in producing the observed asymmetries in detection. In a companion paper, \citet{Hyn2024} analyse the decay phase of SEP events and the possible influence of corotation on this phase of SEP intensity profiles.

In Section \ref{sec.protons_kihara} we describe the main features of the CME dataset and derive distributions of SEP proton event detection.
In Section \ref{sec.messenger} we use the same methodology to derive detection distributions from MESSENGER electron data.
We discuss our results in Section \ref{sec.discuss} and conclusions are summarised in Section \ref{sec.concl}.

\section{Proton events observed by STEREO and GOES}\label{sec.protons_kihara}

\subsection{Dataset of CMEs and associated SEP events}\label{sec.dataset}


In this study, we use the extensive dataset of CME events and associated SEP enhancements gathered by  \citet{Kih2020}. 
They considered all CMEs observed by the Large Angle and Spectrometric COronagraph (LASCO) instrument  \citep{Bru1995} on board the SOlar and Heliospheric Observatory  (SOHO, \citealt{Dom1995}) between December 2006 and October 2017 and selected those with a plane-of-the sky speed $v_{\mathit{CME}}$  greater than 900 km s$^{-1}$ and observed angular width greater than 60$^{\circ}$. Using EUV data from STEREO A, STEREO B and the Solar Dynamics Observatory  (SDO, \citealt{Pes2012}) , for the majority of the CMEs they were able to identify the source AR, both on the far side and front side of the Sun.
Thus they obtained a set of 239 CMEs with information on the source AR location over 360$^{\circ}$ around the Sun and CME properties.

\citet{Kih2020} also analysed in-situ energetic particle  data from STEREO A, STEREO B and GOES to verify whether, for $>$10 MeV  protons,  a flux increase  greater than 1 pfu (particle flux unit, defined as particles s$^{-1}$ sr$^{-1}$ cm$^{-2}$) was associated to CME events in their list, at each of the spacecraft. For GOES they made use of the standard  $>$10 MeV integral channel of the EPS instrument \citep{Ons1996}. For STEREO they combined  LET \citep{Mew2008a} and HET  \citep{vRos2008} channels to obtain $>$10 MeV proton intensities. 
Times when no SEP data were available were excluded, as well as times with high background. Only clear SEP events with unambiguous solar sources were retained.

For 149 of the CMEs in the dataset, three spacecraft were available at different locations in the heliosphere for possible SEP detection.  In some cases no SEP data were available at one or more spacecraft or they were not usable due to contamination by other events. Therefore 48 CMEs had two spacecraft in total available for possible SEP detection and 34 had one.
Overall a set of 577 CME-observer \lq pairs\rq\  was
obtained: for each pair, information was recorded on the spacecraft location with respect to the source AR of the CME and whether or not SEPs were observed. If particles did reach the observing spacecraft, properties of the event such as its onset time, rise time, peak intensity and duration were derived. 
This dataset, detailed in Table 1 of \citet{Kih2020}, forms the basis of our analysis of SEP proton detection asymmetries.
Among the properties of each CME-observer pair, they calculated the  geometrical separation in longitude $\Delta \phi_{\mathit{geom}}$ between source AR and observing spacecraft (termed \lq CME source longitude\rq\ in their plots) given by:
\begin{equation}
\Delta \phi_{\mathit{geom}} = \phi_{\mathit{AR}} - \phi_{\mathit{sc}}
\end{equation}
where $\phi_{\mathit{AR}} $ is the
longitude of the source AR and $\phi_{\mathit{sc}}$ the spacecraft longitude.
The location of the source AR was identified by  \citet{Kih2020} through analysis of data from EUV imagers on both the front side and far side of the Sun.

A very important property of the dataset is that both front-side (from the point of view of Earth) and far-side CME sources were identified. In addition the STEREO spacecraft were magnetically connected to far-side solar longitudes for a large fraction of the time range under study. This means that the dataset does not have a front-side (Earth) bias, neither in the flare observations nor in the SEP observations. 
\citet{Kih2020} presented the  $\Delta \phi_{\mathit{geom}}$ distribution of the 577 pairs (shown in their Figure 1a), showing good coverage of the 360$^{\circ}$ around the Sun in terms of spacecraft locations with respect to the source AR.


 \begin{figure}
   \centering
   \includegraphics[scale=.45]{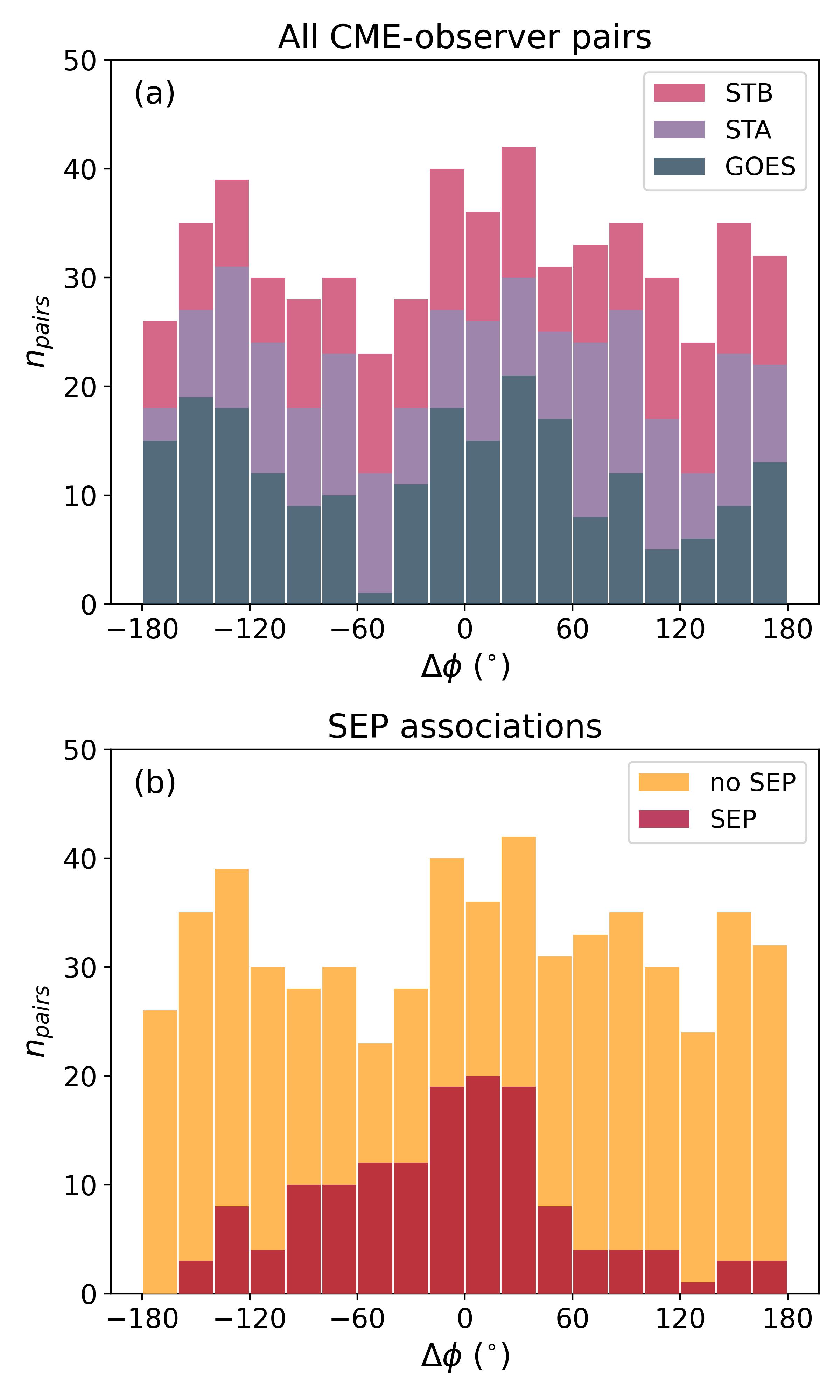}
 \caption{Histogram of CME-observer pairs included in the study of $>$10 MeV proton events, versus $\Delta \phi$.  ({\it a}): stacked plot showing  STEREO A, STEREO B and GOES pairs; ({\it b}): pairs without SEPs at the observer stacked on top of those with SEPs. $\Delta \phi$ was calculated using the actual measured solar wind speed at the spacecraft.}
\label{fig.hist_deltaphi_all}  
     \end{figure}


   \begin{figure}
   \centering
   \includegraphics[scale=.45]{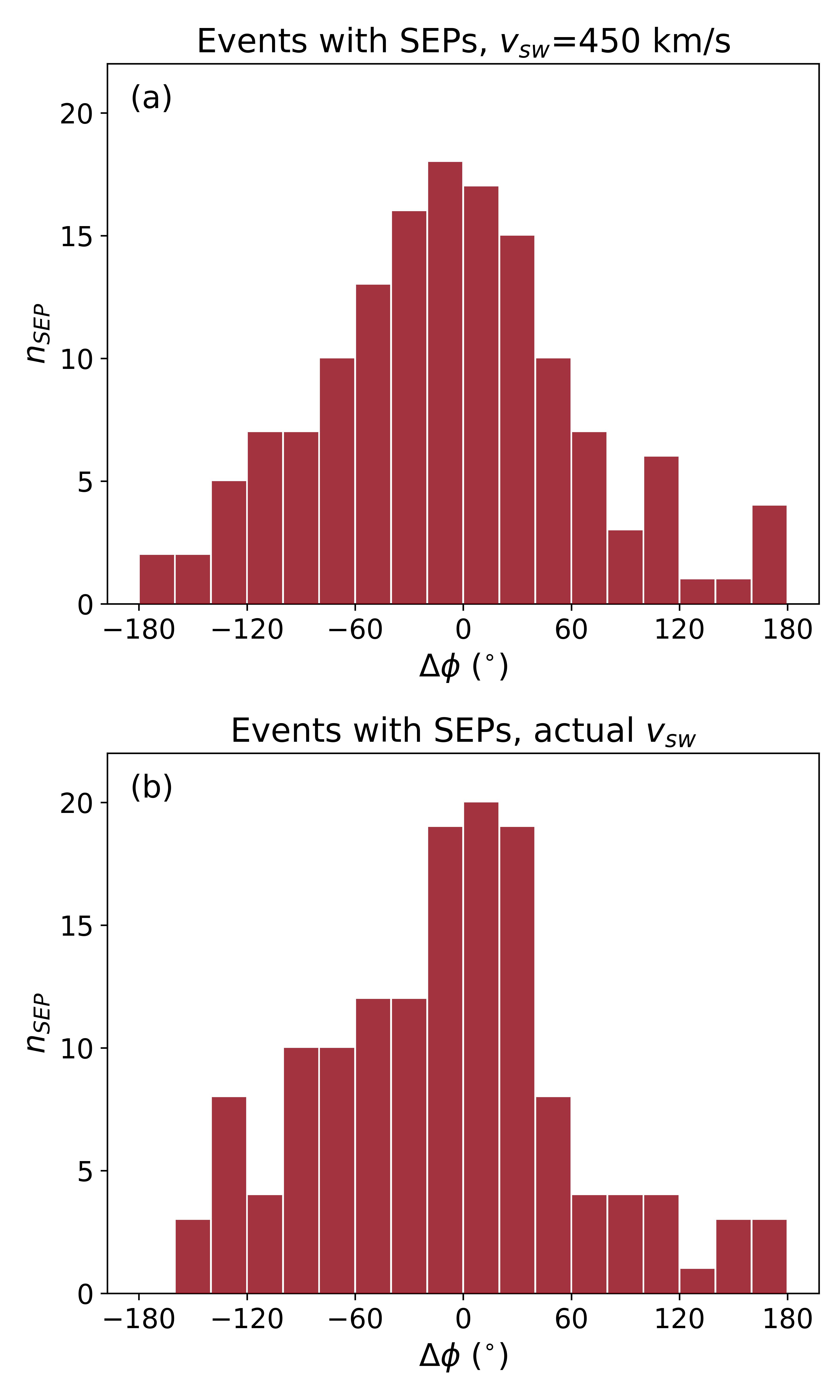}
   \caption{Histogram of CME-observer pairs with SEPs observed, with $\Delta \phi$ calculated using ({\it a}) an assumed $v_{sw}=450$ km s$^{-1}$ and ({\it b})  the actual measured solar wind speed. }
\label{fig.hist_deltaphi}  
     \end{figure}

 In this study we used the data  of  \citet{Kih2020} to calculate
the longitudinal separation $\Delta \phi$ between source AR and observer magnetic footpoint (sometimes termed connection angle) as:
\begin{equation}
\Delta \phi = \phi_{\mathit{AR}} - \phi_{\mathit{ftpt}}  \label{eq.delta_phi}
\end{equation} 
where $\phi_{\mathit{ftpt}}$
is the longitude of the footpoint of the interplanetary magnetic field (IMF) line through the observer. A negative  $\Delta \phi$  indicates a source AR  to the east of the spacecraft footpoint, while a positive  $\Delta \phi$  a source to its west.
We derived $\phi_{\mathit{ftpt}}$ by assuming a Parker spiral IMF between the spacecraft and the Sun, calculated with either
the measured solar wind speed at the spacecraft or, for comparison,  a constant value $v_{\mathit{sw}}$=450 km s$^{-1}$.
We note that $\Delta \phi$ takes different values for two events with the same geometry but different solar wind speed, unlike  $\Delta \phi_{\mathit{geom}}$.

Figure \ref{fig.hist_deltaphi_all} displays the distribution of the CME-observer pairs versus $\Delta \phi$, calculated using the measured solar wind speed at the spacecraft.
In the top panel we show the contribution of the three different spacecraft to the different bins and in the bottom panel we present the two populations of pairs, namely with and without SEP events, including all observing spacecraft.
The $\Delta \phi$ distribution shows excellent spacecraft coverage over the 360$^{\circ}$ range of  $\Delta \phi$ values for the potential detection of SEPs, including a large number of instances where the spacecraft footpoints were located at wide longitudinal separation from the AR.

\subsection{Occurrence distribution of proton SEP events}\label{sec.occur_protons}

The bottom panel of  Figure \ref{fig.hist_deltaphi_all} displays the subset of CME-observer pairs for which an SEP event was observed (144 pairs, {\it red bars}) over the histogram of all pairs (577 pairs, {\it orange bars}).   Here it is evident that even for locations close to the ideal magnetic connection ($\Delta \phi$=0) a significant number of events did not result in SEPs.

Comparing points either side of $\Delta \phi$=0 (best possible magnetic connection to the source AR), an asymmetry is evident, with configurations with negative $\Delta \phi$ more likely to result in the detection of an SEP event compared to those with positive one. This asymmetry is not immediately visible in the related $\Delta \phi_{\mathit{geom}}$ histogram presented by \citet{Kih2020}, though it is present also in that plot (their Figure 1b).

\begin{figure}
   \centering
   \includegraphics[scale=.49]{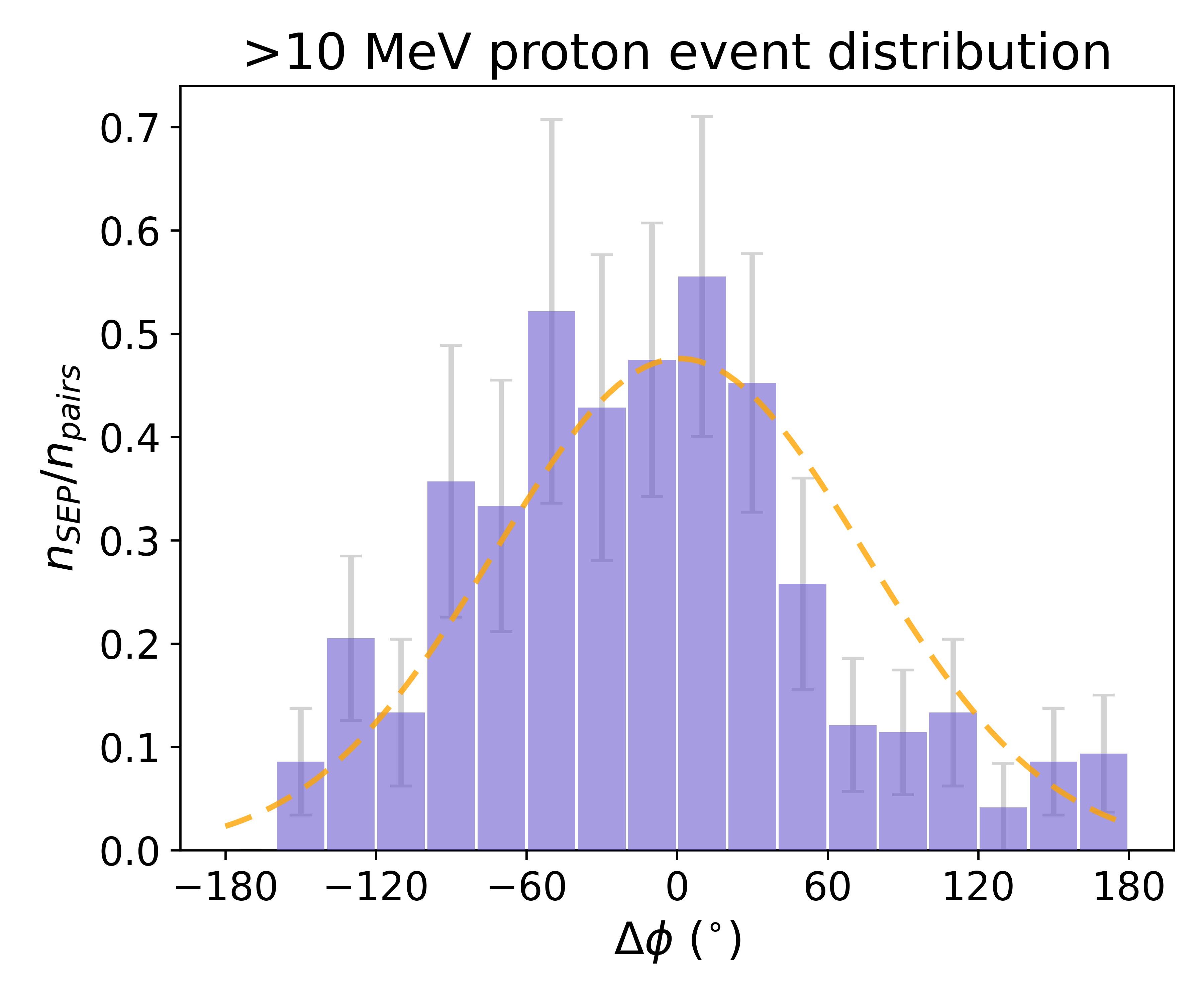}
       \caption{Distribution in $\Delta \phi$  of CME-observer pairs that resulted in SEPs being
observed. Dashed line: gaussian fit to $\Delta \phi$$<$0 portion of the histogram, mirrored to $\Delta \phi$$>$0.}
   \label{fig.distrib}
     \end{figure}

We explore the asymmetry in more detail in Figure \ref{fig.hist_deltaphi}, which presents histograms of pairs for which SEPs were observed, for the case when $\Delta \phi$  is calculated using ({\it a}) a constant  $v_{\mathit{sw}}$=450 km s$^{-1}$  and  ({\it b}) the actual measured solar wind speed.
Comparing the two panels one can see that using the actual $v_{sw}$ measured at the spacecraft produces a histogram that is more peaked around 0 and with a more pronounced lack of events for $\Delta \phi$$>$40$^{\circ}$.

A distribution of SEP detections can be obtained by dividing $n_{\mathit{SEP}}$, the number of pairs with SEPs of  Figure~\ref{fig.hist_deltaphi}b (using the actual $v_{\mathit{sw}}$), by the total number $n_{\mathit{pairs}}$ of CME-observer pairs in each bin (Figure  \ref{fig.hist_deltaphi_all}): the result is shown in  
Figure ~\ref{fig.distrib}. Error bars are calculated by assuming Poisson errors for  $n_{\mathit{SEP}}$ and  $n_{\mathit{pairs}}$ and propagating the error to the ratio.
The yellow dashed line is a gaussian fit to the $\Delta \phi$$<$0 portion of the histogram, mirrored to  $\Delta \phi$$>$0.

Excluding the well-connected longitude range, defined here as the region where $|\Delta \phi|$$<$40$^{\circ}$ (the four central bins in Figure \ref{fig.distrib}),
there is a  strong asymmetry in the $\Delta \phi$ distribution of SEP event occurrence  with respect to $\Delta \phi$=0, with an excess of events in the negative side and a lack of events on the positive one.   By summing $n_{\mathit{SEP}}/n_{\mathit{pairs}}$ for  bins outside the well connected region on each side of the histogram, we find that observers for which the source AR  is on the east side of spacecraft footpoint  are 93\% more likely to detect an SEP event compared to observers with source AR on the west side.
The mean of the distribution is $\overline{\Delta\phi}=-12^\circ$ and its standard deviation is $\sigma_{\Delta\phi}=72^\circ$. 

We tested the asymmetry of the SEP distribution in Figure  \ref{fig.distrib}  by using a sign test. The sign test can be used to evaluate the null hypothesis that the distribution is symmetric with respect to a given $\Delta\phi_0$, without making any assumptions on the shape of the distribution. Under the null hypothesis, the numbers of events with $\Delta\phi<\Delta\phi_0$ and $\Delta\phi>\Delta\phi_0$
follow a binomial distribution with 50\% probability that an event is at either side of  $\Delta\phi_0$.
The statistical significance of the hypothesis can be evaluated with a standard binomial test. The sign test requires as input the number of events whereas the distribution in Figure  \ref{fig.distrib} represents the number of events per pair. We obtained the number of events  by multiplying the values of $n_{sep}/n_{pairs}$  by the mean number of pairs per bin, obtained by averaging the histogram of Figure \ref{fig.hist_deltaphi_all}. We used a one-sided sign test to assess whether the distribution in Figure  \ref{fig.distrib}  is symmetric with respect to $\Delta\phi_0=0$ and found that the null hypothesis can be rejected with a p-value of 0.038, showing that there is only a 3.8\% probability that the null hypothesis is correct.
Thus the test implies there that the underlying distribution is asymmetric with respect to 0, with an excess of events for $\Delta\phi$$<$0 compared to the  $\Delta\phi$$>$0 side.
Based on the sign test we conclude that the asymmetry  with respect to $\Delta\phi=0$  is statistically significant and
SEP events are much more likely to be observed
when the source AR is in the east with respect to the observer footpoint.

We further investigated the shape of the distribution. We calculated its Pearson-Fisher skewness parameter $S$, obtaining a value $S$=0.40 (for comparison, the skewness of the half-normal distribution is 1).
  We compared the distribution to a normal distribution with the same mean $\overline{\Delta\phi}$ and sample standard deviation $\sigma_{\Delta\phi}$ using a $\chi^2$ test, normalising both distributions to the representative event numbers using the same approach as we did for the sign test. We further summed bins where the representative count in the normalised normal distribution was smaller than~5, as small counts are known to give erroneous results within the $\chi^2$ test. We found that the null hypothesis of normality of the distribution could not be rejected (p-value 0.63).
  Therefore the observed distribution is statistically compatible with an underlying gaussian distribution peaking at $\overline{\Delta\phi}=-12^\circ$.
  Finally, we applied the sign test with $ \Delta\phi_0=-12^\circ$ and found that
  the hypothesis of symmetry of the distribution with respect to the observed mean cannot be rejected  (p-value 0.78).
We note that the results of the statistical tests outlined above vary slightly depending on the exact $\Delta\phi$ binning used, however the conclusions remain unchanged irrespective of binning.

\section{Electron events observed by MESSENGER}\label{sec.messenger}

\citet{Rod2023b} carried out an extensive analysis of 61 solar energetic electron events detected at the MESSENGER spacecraft between 2010 and 2015. Events were identified using the 71-112 keV electron channel of the EPS instrument, part of the EPPS suite \citep{And2007}. During the events, the spacecraft was located at heliocentric distances between 0.31 and 0.47 au. Of the 61 events, 57 were associated with a CMEs, as detailed in \citet{Rod2023a}.

\begin{figure}
\centering
\includegraphics[scale=.45]{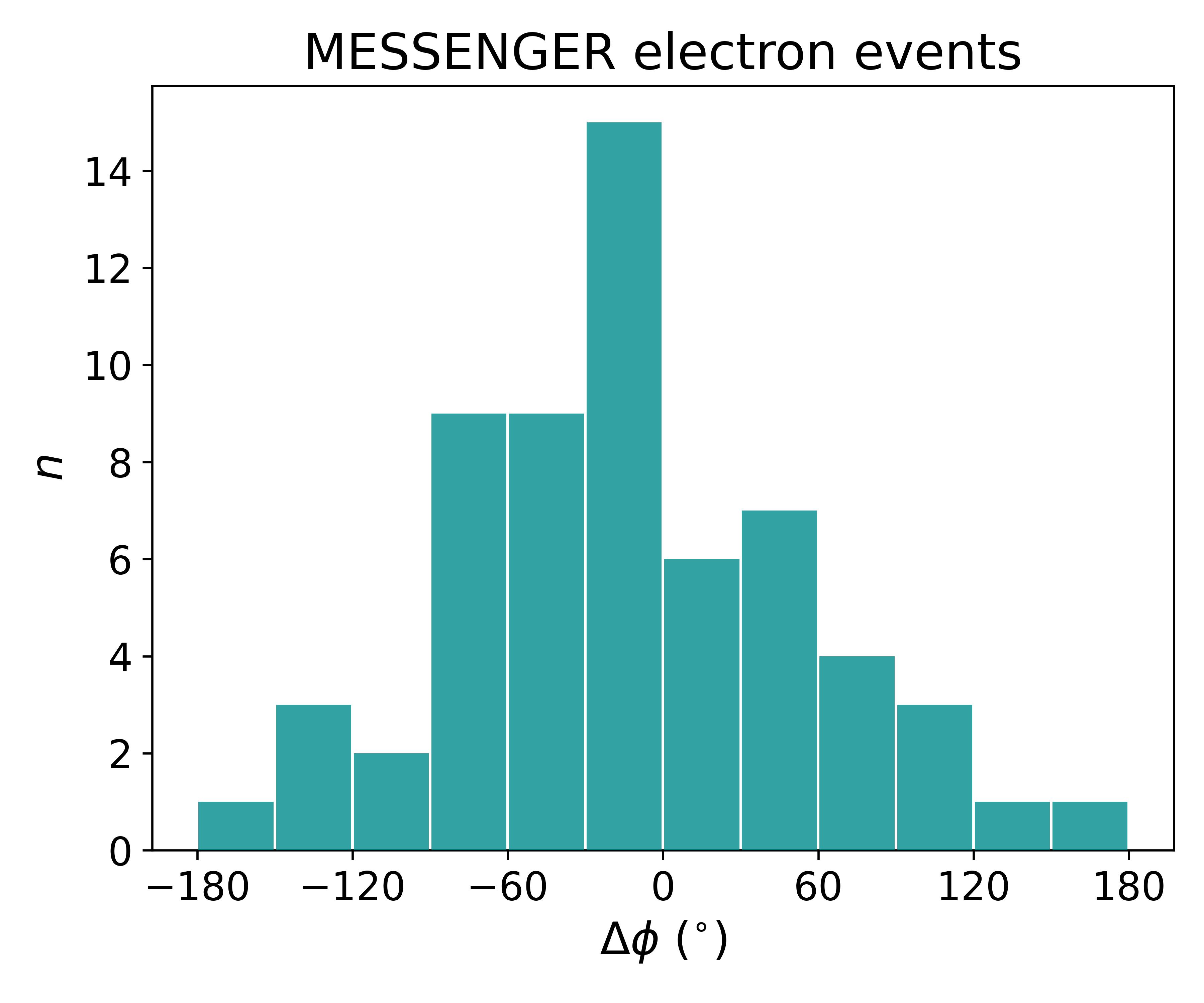}
\caption{Histogram of  $\Delta \phi$  values for the 61 MESSENGER electron events studied by \citet{Rod2023b}. The  71-112 keV electron channel was used to identify events.  }
\label{fig.hist_deltaphi_mess_initial}  
\end{figure}

Considering  Figure 1a of \cite{Rod2023b},  displaying electron event peak intensity versus  $\Delta \phi$ (which they term connection angle, CA,  defined as in Eq.(\ref{eq.delta_phi})), an E-W asymmetry in detection is visible in their plot. The asymmetry is explored further in Figure 
\ref{fig.hist_deltaphi_mess_initial}, displaying the histogram of number of events versus connection angle. Solar wind speed measurements are not available for MESSENGER therefore for the calculation of the spacecraft footpoint a solar wind speed  $v_{sw}$=400 km s$^{-1}$ was assumed, and the longitude of the source AR is that of the flare associated to the event \citep{Rod2023a}.
An asymmetry similar to that shown in Figure \ref{fig.hist_deltaphi} for protons can be seen.  The total number of SEP events at MESSENGER is smaller than that of the proton events of Figure \ref{fig.hist_deltaphi}, due to the high background of the MESSENGER particle instrument, as discussed in  \citet{Rod2023b}, and the shorter time range.

  \begin{figure}
   \centering
   \includegraphics[scale=.45]{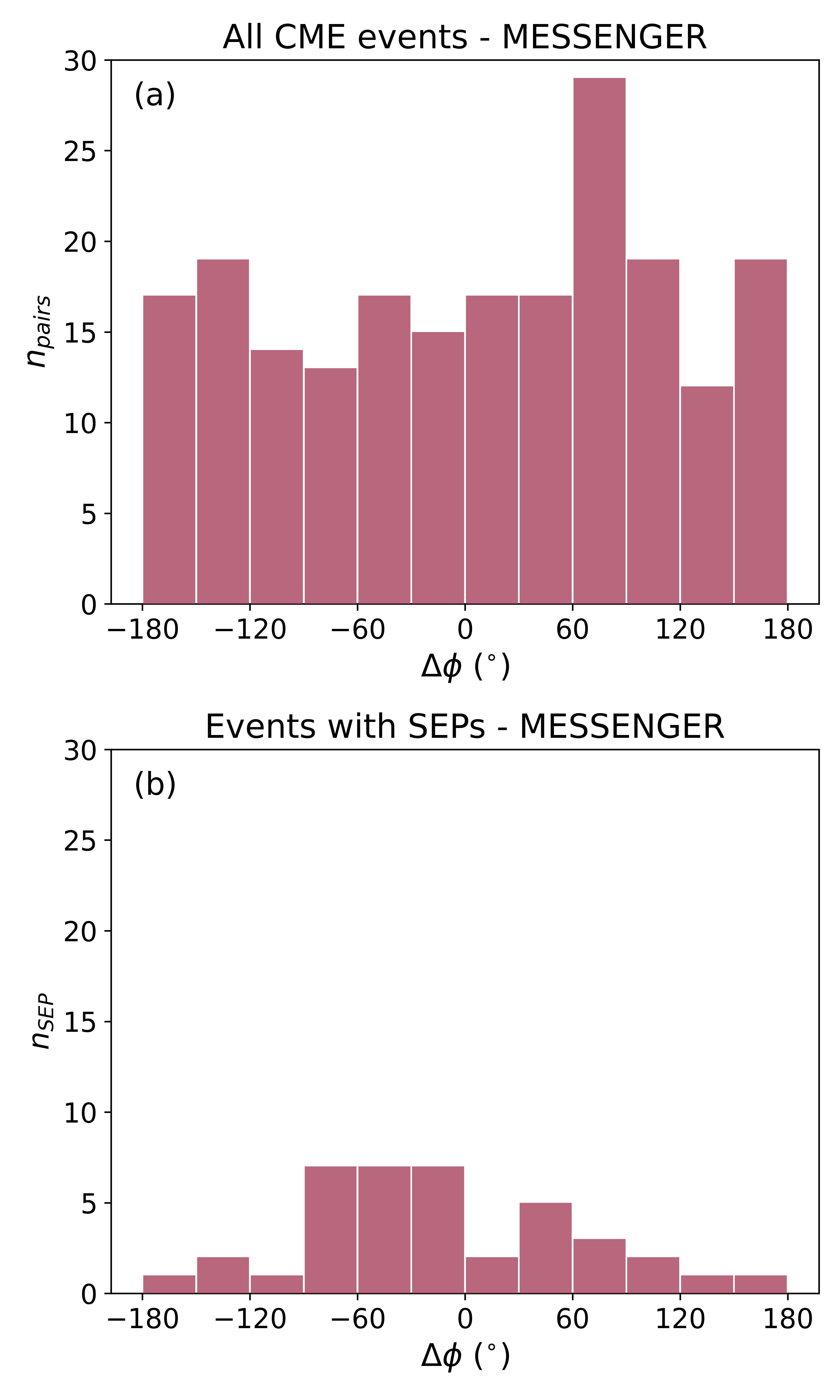}
   \caption{Histogram of  $\Delta \phi$  values for ({\it a})  all CME events in the Kihara et al.~(2020) dataset for which MESSENGER SEP observations were available and  ({\it b}) subset of events in  ({\it a}) for which SEP electrons were observed (39 events). }
\label{fig.hist_deltaphi_mess}  
     \end{figure}

\subsection{Occurrence distribution of electron SEP events}\label{sec.occur_electrons}

We derived an occurrence distribution of electron events  versus  $\Delta \phi$ based on the dataset of \citet{Rod2023b} using a methodology similar to that of Section \ref{sec.occur_protons}. Starting from the list of 239 CMEs of \cite{Kih2020},
we verified whether MESSENGER electron observations were available, this being the case for 208 CMEs, which we used for our analysis.
For each one we calculated the value of $\Delta \phi$ assuming a Parker spiral IMF. While the CMEs used in this analysis are the same as in Section \ref{sec.protons_kihara}, the $\Delta \phi$ values for MESSENGER are different from those of the STEREO and GOES spacecraft.  Figure \ref{fig.hist_deltaphi_mess}a shows the distribution of $\Delta \phi$ values for the CME-MESSENGER pairs.
As was the case for the analysis in Section \ref{sec.protons_kihara}, there is fairly uniform coverage of the 360$^{\circ}$ around the Sun. We note that  the bin centered at   $\Delta \phi$=75$^{\circ}$ shows a much larger number of CME events compared to the other bins.

We then analysed whether an electron event took place at MESSENGER in association with the CME event. Figure \ref{fig.hist_deltaphi_mess}b shows the distribution in $\Delta \phi$ of the events for which an SEP enhancement was detected.  The number of events in this histogram is smaller than that of the histogram of Figure \ref{fig.hist_deltaphi_mess_initial} because not all the 61 SEP events of Figure \ref{fig.hist_deltaphi_mess_initial} have an associated CME that meets the selection criteria for inclusion in  the \cite{Kih2020} study (i.e.~ some have an associated CME with speed smaller than 900 km s$^{-1}$ and angular width smaller than 60$^{\circ}$).

\begin{figure}
   \centering
   \includegraphics[scale=.49]{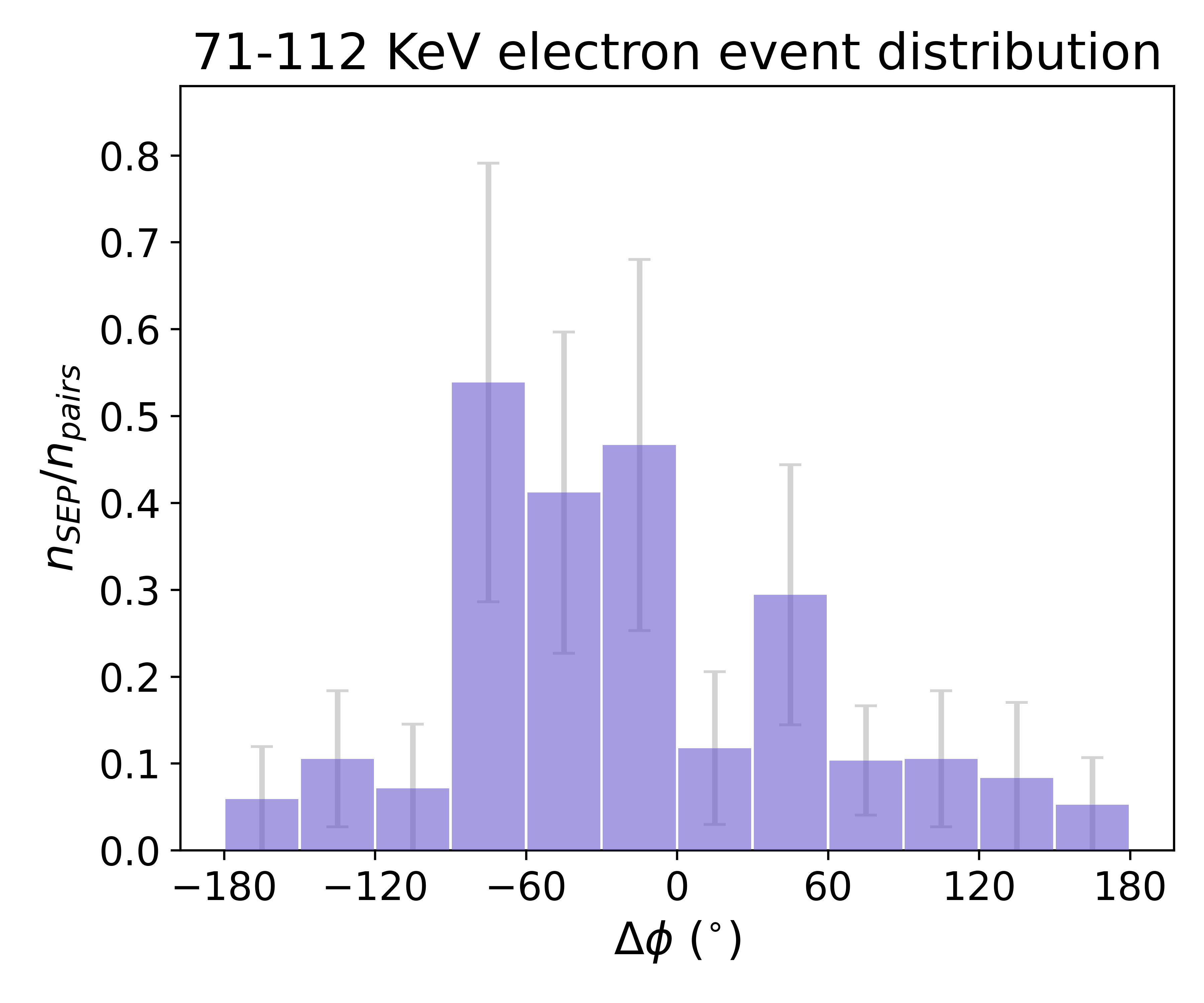}
       \caption{Distribution in $\Delta \phi$  of CME events that resulted in SEP electrons being
observed at MESSENGER.}
   \label{fig.distrib_mess}
     \end{figure}

By dividing the histogram of Figure \ref{fig.hist_deltaphi_mess}b by that of Figure \ref{fig.hist_deltaphi_mess}a, the distribution of events with SEPs shown in 
Figure \ref{fig.distrib_mess} was obtained.
The total number of events is much smaller compared to the case of Figure \ref{fig.distrib}, due to the shorter time range over which MESSENGER data are available and the higher instrumental background. However there are indications that the same detection asymmetry seen in the solar energetic proton event distribution is present also for electron events. It should be noted that despite the much larger number of CME events in the  $\Delta \phi$=75$^{\circ}$ bin compared to all the other bins, $n_{\mathit{SEP}}$ in this bin is low.
Considering only configurations with $|\Delta \phi|$$>$30$^{\circ}$,  events with negative $\Delta \phi$ are 86\% more likely than those with positive $\Delta \phi$. The mean of the distribution is $\overline{\Delta\phi}=-18^\circ$ and its standard deviation is $\sigma_{\Delta\phi}$=74$^\circ$.

Applying the sign test in a similar way as for the proton data, we found that the distribution in Figure  \ref{fig.distrib_mess}  is asymmetric with respect to $\Delta\phi=0$  with p-value of 0.017.
  The null hypothesis (underlying distribution symmetric with respect to $\Delta\phi$=0) has a probability of 1.7\% and is thus rejected.
  Therefore the detection asymmetry is present also in the electron data.
  The Pearson-Fisher skewness parameter for the distribution is $S$=0.52. We could not perform a test of gaussianity as the number of events is not sufficient for the $\chi^2$ test. Applying the sign test with respect to $\Delta\phi_0=-18^\circ$,
   the assumption of symmetry with respect to the mean cannot be rejected (p-value 0.62).



 \section{Discussion} \label{sec.discuss}

Indications that an E-W asymmetry in SEP event detection with respect to the best possible magnetic connection ($\Delta \phi$=0) may be present in the data have been discussed in the literature in the past.
\citet{Kal1992b} presented an analysis of 77 SEP events detected by the Helios 1 and 2 spacecraft between 1974 and 1985. This dataset was affected by Earth bias in the flare observations but not the SEP observations. Their Figure 4 displayed the distribution of the events in the $\Delta \phi$-heliolatitude plane: they noted an asymmetry in the east-west distribution of the events, with $\sim$2/3 lying in the $\Delta \phi$$<$0 portion of the plane. They stated: \lq We think that this asymmetry has no physical reason but rather suspect that it is due to the Helios orbit\rq. They argued that events with large positive $\Delta \phi$ would require the Helios spacecraft to be located behind the east limb and thus proposed that they may have been less numerous due to poor data transmission or radio blackouts \citep{Kal1992b}.

\cite{Dal2003} analysed 52 of the same SEP events (the subset of events identified as gradual) using data from Helios 1, Helios 2 and IMP8: they used 4--10 MeV and 28--36 MeV proton channels and the 0.7--2.0 MeV electron channel and plotted the duration of the SEP events versus $\Delta \phi$. The resulting graphs showed that events with large negative  $\Delta \phi$ tended to have the longest durations and those with large positive  $\Delta \phi$ had much shorter duration. They also considered the E-W asymmetry in event occurrence commented upon by \citet{Kal1992b}  and carried out an analysis of the trajectories of the Helios 1 and 2 spacecraft: this showed that their orbits did not make events with  large positive $\Delta \phi$ less likely. They also pointed out that spacecraft-AR configurations leading to large positive  $\Delta \phi$ were possible that did not involve the spacecraft being at risk of data transmission problems. Thus they termed this east-west asymmetry the \lq detection asymmetry\rq\ and argued that it is a real physical effect.

 \citet{He&Wan_2017} used  a list of 78 major solar proton events at Earth during 1996--2011,  produced by the NOAA Space Weather Prediction Centre, to show that, for $|\Delta \phi|$$\lesssim$40$^{\circ}$, there is an excess of events for negative $\Delta \phi$ values compared to positive ones. This study is affected by Earth bias in flare observations since only observations of SEP source regions on the front side of the Sun were available in compiling the list (with a small fraction of the SEP events having been associated to regions that rotated over the west limb of the Sun).  It is also affected by Earth bias in SEP observations as only near-Earth data was used. Thus the study could not probe the 40$^{\circ}$$<$$\Delta \phi$$<$180$^{\circ}$ region. \citet{He&Wan_2017} ascribed the asymmetry to perpendicular diffusion effects.

 The results presented in Section \ref{sec.protons_kihara} are based on a much more extensive dataset compared to previous work and use, for the first time in this type of study, consistent information on solar sources of SEP events located on the far side of the Sun. In addition, by starting the analysis from a series of CME events, regardless of whether or not an SEP event was produced, we have been able to derive  distributions of occurrence of SEP events (Figures \ref{fig.distrib} and \ref{fig.distrib_mess}).
The proton distribution shows a clear E-W asymmetry in detection, confirming the earlier findings.
The number of electron events from the MESSENGER dataset discussed in Section \ref{sec.messenger} is smaller than for protons, but a statistically significant asymmetry is present.

One limitation of our study is that it uses a high threshold to define an SEP proton event, because of the reliance on data from the GOES spacecraft, which has a high background. The  MESSENGER electron instrument we utilised also has a high background so that again a high threshold was employed in the detection of electron events. Thus overall the study has an emphasis on intense SEP events. The CME sample is focussed on fast and wide CMEs. It is hoped that future work will use different selection criteria and extend this type of study to other types of SEP and solar events.

Combining our proton and electron results  with  earlier indications of a similar asymmetry from data from the Helios 1 and Helios 2 spacecraft during 1974--1985 \citep{Kal1992b,Dal2003}, there are three independent datasets probing large positive  $\Delta \phi$ values which confirm that  the detection asymmetry in SEP events is a real effect. Our study shows that in the datasets we analysed the asymmetry arises despite the distribution of solar events with potential to produce SEPs being uniform in  $\Delta \phi$. Thus we conclude that the asymmetry in SEP event occurrence is real and it is caused by a physical effect.

One possible explanation for the E-W asymmetry in SEP detection is that it may be related to corotation of particle-filled magnetic flux tubes with the Sun.  3D test particle simulations have shown that corotation plays an important role in determining the characteristics of SEP events \citep{Mar2015,Dal2017a,Hut2023a}.
For an observing spacecraft for which the source AR is to the east of the magnetic footpoint  ($\Delta \phi$$<$0) corotation sweeps the magnetic field lines towards the observer, while if the source AR is to the west ($\Delta \phi$$>$0), corotation takes magnetic flux tubes away from the observer. In the frame of reference of a magnetic flux tube corotating with the Sun, the particle flux evolves due to: a) time dependent injection of accelerated particles directly into this tube and b) transport of particles in/out of the flux tube as a result of 3D effects. The latter may include perpendicular transport associated with turbulence (e.g.~\citealt{Str2017a,Lai2023}), guiding centre drifts \citep{Dal2013} and heliospheric current sheet drift \citep{Wat2022}.
The SEP flux measured by an observer is the combined effect of time dependent changes due to a) and b) within each flux tube, convolved with the spatial effect associated with corotation of the magnetic flux tubes themselves over the observer.
Since detection of an SEP event requires the intensity to exceed the instrumental background of a given spacecraft detector  (or, as in the present analysis, to cross a specified threshold), when the observer is not directly connected to the source region of the event it is reasonable to assume that, for $\Delta \phi$$>$0 configurations, corotation works against intensities going above background at the observer, while for $\Delta \phi$$<$0 it makes detection more likely by carrying magnetic flux tubes towards the observer. Hence corotation may be a contributing factor to the observed East-West detection asymmetry.
Because of the large variation in magnitude and spatial extent of SEP events it is still possible for events with large positive $\Delta \phi$ to be detected, however these are statistically less likely.

 \begin{figure}
   \centering
   \includegraphics[scale=.49]{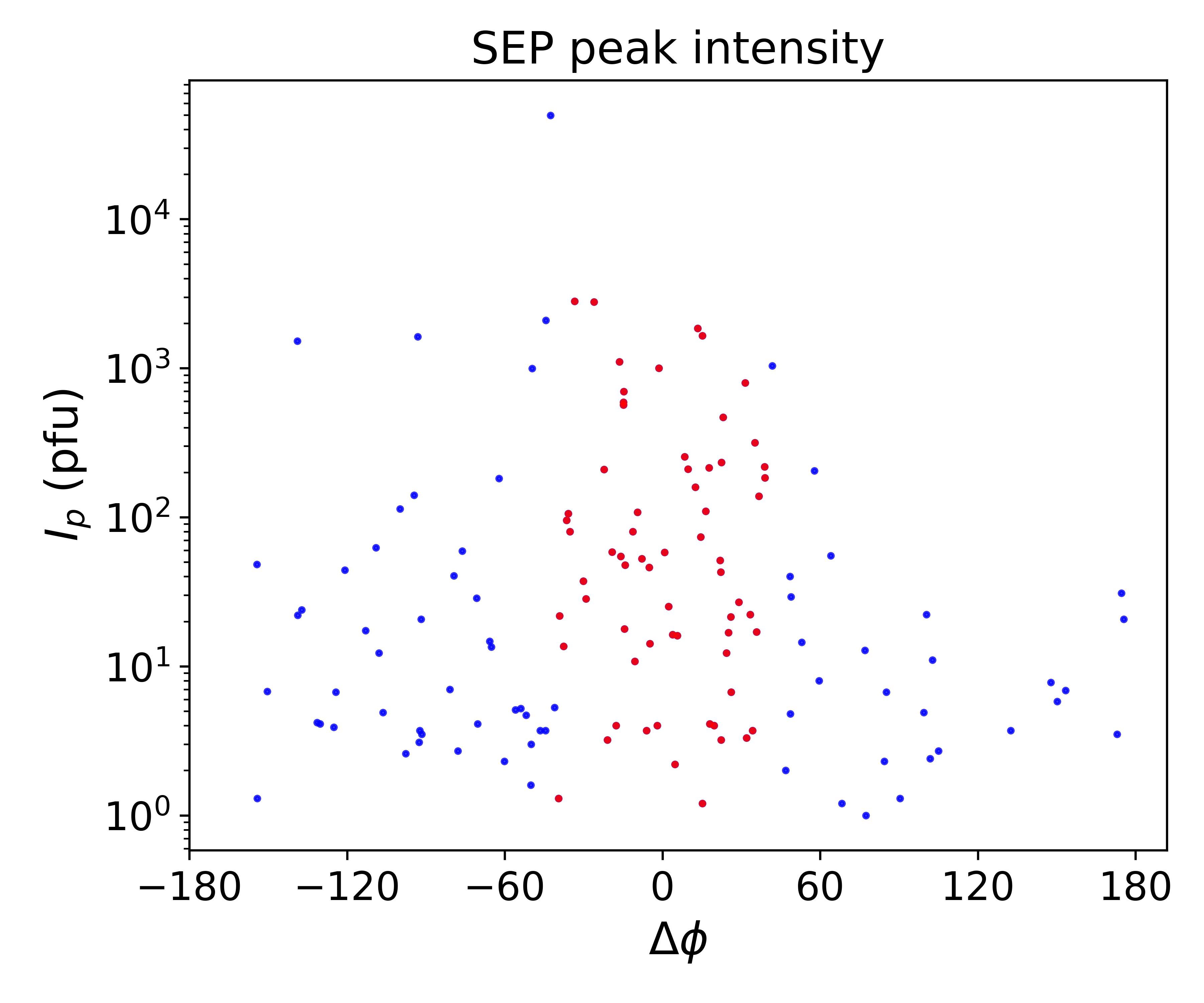}
       \caption{SEP peak intensity  versus  $\Delta \phi$ for the $>$10 MeV proton events. Events with $|\Delta \phi|$$<$40$^{\circ}$ are colour coded in red.}
   \label{fig.peak_intens}
     \end{figure}

As a possible alternative explanation the observed asymmetry may be caused by a systematic E-W variation of
efficiency of SEP energisation along an accelerating shock front  \citep{Can1988a,Tyl2005,Kah2016}. Over time, a given observer is connected to different portions of a propagating CME-driven shock. Following on from ideas proposed by \citet{Sar1984}, \citet{Tyl2005} suggested that given that in part of its front the shock is quasi-parallel while in others it is  quasi-perpendicular, the different efficiencies of acceleration for these two types of shock may explain differences in intensity profile parameters and composition.
As noted by \citet{Kah2016} close to the Sun at the flanks of a CME shock there are no significant differences in shock obliquity between east and west. However as the shock propagates further from the corona an observer with $\Delta \phi$$>$0 will be connected to a quasi-parallel shock, while one with  $\Delta \phi$$<$0  to a quasi-perpendicular one. Thus if what influences detection is acceleration far from the corona and acceleration at quasi-perpendicular shocks is more efficient, this may explain an asymmetry in detection.
It should be noted that the above interpretation assumes a geometry of the CME shock where its flanks curve back towards the Sun, resulting in the quasi-parallel and quasi-perpendicular configurations, initially invoked by \citet{Sar1984} (see their Figure 11) to explain features of SEP events.
However a study by \citet{Can1988b} argued that over an extent of approximately 120 degrees centered at the nose, the expansion speeds of a shock front tend to be similar, resulting in a shock geometry that is semi-circular. In this scenario the angle between shock normal and magnetic field does not vary strongly and the explanation above no longer holds. 
\citet{Din2022} used a 2D model of SEP acceleration at a propagating CME shock to show that even  a symmetric acceleration efficiency at the shock will result in an asymmetric injection with time since the shock connects to westward longitudes as it reaches larger radial distances.

A third possible explanation  is that some asymmetry may be introduced by  perpendicular diffusion processes during the transport of SEPs in the heliosphere  \citep{He&Wan_2017}. \citet{Str2017a} pointed out that the nature of diffusion perpendicular to the Parker spiral interplanetary magnetic field produces a spatial distribution of SEP intensities in the heliosphere with peak located to the west of the best magnetic connection (see their Figure 3), thus producing larger intensities for negative $\Delta \phi$  (using the definition in Eq.(\ref{eq.delta_phi})). The same feature can be observed in the spatial distributions from the model of \citet{Lai2023} (see their Figure 2).
It is also possible that a combination of corotation, asymmetric SEP injection at a CME shock and interplanetary transport processes may contribute to the observed detection asymmetry.

The asymmetry in detection may be related to features in the distribution of SEP peak intensities that have been reported in the literature.
  \citet{Ric2014}  analysed individual three-spacecraft events by fitting a gaussian to a plot of SEP peak intensity versus $\Delta \phi$ at the three spacecraft. They found that there is a tendency for the location,  $\Phi_0$, of the peak of the fitted gaussian to be shifted towards negative  $\Delta \phi$  values (for the definition of $\Delta \phi$ given in Eq.~(\ref{eq.delta_phi})). For 14--24 MeV protons they obtained   $\Phi_0$=$-$15$\pm$35$^{\circ}$.
  In an earlier study \citet{Lar2013} obtained $\Phi_0$ with a different methodology that uses simultaneous fitting of multiple events within the assumption that $\Phi_0$ and the width of the gaussian are the same for all events.
   \citet{Rod2023b} plotted electron peak intensities in the 71-112 keV channel versus  $\Delta \phi$ in their Figure 1a and noted that peak intensities tended to be larger for negative $\Delta \phi$ values. 

Figure \ref{fig.peak_intens} shows SEP peak intensity $I_p$ versus  $\Delta \phi$ for  the $>$10 MeV proton event dataset of Section \ref{sec.protons_kihara}: here one can see that outside of the well connected range, $I_p$ tends to be larger for the negative $\Delta \phi$ range compared to the positive one. Therefore the trend displayed in Figure \ref{fig.peak_intens} agrees with that reported by \citet{Rod2023b} for electrons.
  Given that event detection requires intensities to exceed a threshold (determined either by the instrumental background or the galactic cosmic ray intensity), if there is a tendency for events to be more intense in the negative $\Delta \phi$   range compared to the positive one, this will produce a detection asymmetry such as the one we observed.
  In addition if the  $I_p$ versus  $\Delta \phi$ plot is as shown in  Figure  \ref{fig.peak_intens} and reported in \citet{Rod2023b},  one would expect that for many three-spacecraft events the largest peak intensity is located at negative $\Delta \phi$ values, resulting in the tendency for negative $\Phi_0$ values observed \citep{Lar2013}. Thus the peak intensity trend observed causes the negative $\Phi_0$  values in three-spacecraft gaussian fits.
Therefore we conclude that the underlying physics is likely the same for the detection asymmetry, the peak intensity asymmetry and the negative $\Phi_0$ gaussian fit results.
Corotation causes peak intensities  to be larger for negative  $\Delta \phi$ because particle filled flux tubes corotate towards the observer, as shown by \citet{Hut2023a}, and thus can explain the asymmetry in peak intensities.
\citet{Str2017a} ascribed the negative $\Phi_0$ gaussian fit results to perpendicular diffusion, within a model that did not include corotation effects. 
  Alternatively variation of acceleration efficiency along the shock could be causing the observed peak intensity trend.

Regarding the fact that both $>$10 MeV proton and 71-112 keV electron data show indications of an east-west asymmetry in detection,
it should be noted that the sources of SEP electrons and protons may be different and considerable debate exists in the literature on this point.  Some electron events may have origin in localised regions, including flare reconnection regions.
Within the corotation interpretation however, whatever the origin of the energetic particles, the magnetic flux tubes in which they are injected will be subject to rotation of the magnetic field lines and thus an asymmetry would be expected for both electrons and protons.

Test particle simulations of 5 MeV protons injected by a wide propagating shock-like source showed that, within an individual event, corotation results in a long-duration SEP decay phase at an observer with $\Delta \phi$$<$0 (eastern AR with respect to the observer's magnetic footpoint), while it contributes to the SEP event being \lq cutoff\rq\ for cases with  $\Delta \phi$$>$0 (western AR with respect to the footpoint) \citep{Hut2023a}. They also demonstrated that, once corotation is included,  the decay time constant of the event is independent of the scattering mean free path.

In a recent study \cite{Hyn2024} analysed the decay phase of SEP events and showed that within individual events the decay time constant displays a systematic decrease with  $\Delta \phi$: according to test particle simulations this systematic behaviour is the result of corotation effects  \citep{Hut2023a}. Thus both detection asymmetry and analysis of SEP decay phases appear to point towards corotation playing an important role.



  \section{Conclusions} \label{sec.concl}

 In this paper we analysed the distribution of occurrence of SEP events with respect to the connection angle $\Delta \phi$, the longitude separation between source AR location and observer magnetic footpoint (Eq.~(\ref{eq.delta_phi})), using datasets that do not suffer from Earth bias.

Our main conclusions are as follows:

\begin{itemize}

\item
  Based on a dataset of 577 CME-observer pairs during 2006--2017 \citep{Kih2020}  the distribution of occurrences of  $>$10 MeV proton SEP events at 1 au displays an east-west asymmetry with respect to $\Delta \phi$=0. Outside the well-connected longitude range, i.e.~for $|\Delta \phi|$$>$40$^{\circ}$, events with negative $\Delta \phi$ are 93\% more likely than those with positive $\Delta \phi$. Based on a sign test the asymmetry is statistically significant with the null hypothesis (no asymmetry with respect to $\Delta \phi$=0) having probability of 3.8\% and thus being rejected.
  

 \item
   Occurrences of 71-112 keV electron SEP events measured by MESSENGER at radial distances between 0.31 and 0.47 au during 2010--2015 \citep{Rod2023b} also display a similar asymmetry, with a higher likelihood of events for negative  $\Delta \phi$,  though with lower total event numbers.   A sign test rejects the null hypothesis (no asymmetry  with respect to $\Delta \phi$=0) since it has a probability of 1.7\%.

\end{itemize}

As for $\Delta \phi$$<$0 corotation sweeps particle-filled magnetic flux tubes towards the observer and for  $\Delta \phi$$>$0 it moves them away from the observer, corotation as a spatial effect can provide an explanation to the observed asymmetry. Other effects such as east-west differences in the efficiency of acceleration at a CME driven shock front  or perpendicular transport effects may be considered as alternative explanations.
 It is interesting that the effect appears to be present for both electrons and protons, and is seen both close to the Sun (0.31--0.47 au) and at 1 au. The corotation explanation can account for these observations, as it acts in the same way on protons and electrons and it is present at both radial distances.  It is hoped that future modelling will explore whether asymmetric acceleration at a CME shock or perpendicular transport can reproduce the observed features at various radial distances and for different species.

The distribution of event detections can be described as either characterised by an asymmetry with respect to $\Delta \phi$=0 or as displaying a shift of its peak towards negative $\Delta \phi$ values, with the mean of the distribution being located at  $\overline{\Delta\phi}=-12^\circ$ for $>$10 MeV protons and $\overline{\Delta\phi}=-18^\circ$ for 71-112 keV  electrons. It is likely that the physical mechanism responsible for these features is the same as for the asymmetry in peak intensity distribution \citep{Rod2023b} and offset in gaussian fit peaks  \citep{Lar2013,Ric2014}. 






Finally, if events with large positive $\Delta \phi$  are much less likely, as our study indicates, this has potential consequences for Space Weather in terms of developing empirical tools and methodologies for SEP forecasting.


\begin{acknowledgements}
  SD, TL and CW  acknowledge support from the UK STFC (grants ST/V000934/1 and ST/Y002725/1) and NERC (via the SWARM project, part of the SWIMMR programme, grant NE/V002864/1). RH acknowledges support from a Moses Holden studentship. AH would like to acknowledge support from from STFC via a doctoral training grant, the University of Maryland Baltimore County (UMBC), the Partnership for Heliophysics and Space Environment Research (PHaSER), and NASA/GSFC. NVN has been supported by NASA grant 80NSSC24K0175. L.R.-G.\ acknowledges support through the European Space Agency (ESA) research fellowship programme. CW's research is supported by the NASA Living with a Star Jack Eddy Postdoctoral Fellowship Program, administered by UCAR’s Cooperative Programs for the Advancement of Earth System Science (CPAESS) under award \#80NSSC22M0097[1].\\
 {\bf Data Access Statement}: The datasets used in this paper are publicly available as follows. The CME event and SEP proton enhancement dataset used in Section \ref{sec.protons_kihara} has been published in \cite{Kih2020}, with their Table 1 providing all the relevant data. The electron SEP event dataset used for the analysis of Section \ref{sec.messenger} has been published in \cite{Rod2023b} in their Appendix A, Table A.1.
  \end{acknowledgements}

\bibliographystyle{aa} 
\bibliography{corot_biblio} 

\begin{thebibliography}{51}
\expandafter\ifx\csname natexlab\endcsname\relax\def\natexlab#1{#1}\fi

\bibitem[{{Andrews} {et~al.}(2007){Andrews}, {Zurbuchen}, {Mauk}, {Malcom},
  {Fisk}, {Gloeckler}, {Ho}, {Kelley}, {Koehn}, {Lefevere}, {Livi}, {Lundgren},
  \& {Raines}}]{And2007}
{Andrews}, G.~B., {Zurbuchen}, T.~H., {Mauk}, B.~H., {et~al.} 2007, \ssr, 131,
  523

\bibitem[{Brueckner {et~al.}(1995)Brueckner, Howard, Koomen, Korendyke,
  Michels, Moses, Socker, Dere, Lamy, Llebaria, Bout, Schwenn, Simnett,
  Bedford, \& Eyles}]{Bru1995}
Brueckner, G.~E., Howard, R.~A., Koomen, M.~J., {et~al.} 1995, Solar Physics,
  162, 357

\bibitem[{{Burlaga}(1967)}]{Bur1967}
{Burlaga}, L.~F. 1967, \jgr, 72, 4449

\bibitem[{{Cane}(1988)}]{Can1988b}
{Cane}, H.~V. 1988, \jgr, 93, 1

\bibitem[{{Cane} {et~al.}(1988){Cane}, {Reames}, \& {von
  Rosenvinge}}]{Can1988a}
{Cane}, H.~V., {Reames}, D.~V., \& {von Rosenvinge}, T.~T. 1988, \jgr, 93, 9555

\bibitem[{{Cohen} {et~al.}(2021){Cohen}, {Li}, {Mason}, {Shih}, \&
  {Wang}}]{Coh2021}
{Cohen}, C. M.~S., {Li}, G., {Mason}, G.~M., {Shih}, A.~Y., \& {Wang}, L. 2021,
  in Solar Physics and Solar Wind, ed. N.~E. {Raouafi} \& A.~{Vourlidas},
  Vol.~1, 133

\bibitem[{{Cohen} {et~al.}(2017){Cohen}, {Mason}, \& {Mewaldt}}]{Coh2017}
{Cohen}, C.~M.~S., {Mason}, G.~M., \& {Mewaldt}, R.~A. 2017, \apj, 843, 132

\bibitem[{{Dalla}(2003)}]{Dal2003}
{Dalla}, S. 2003, in American Institute of Physics Conference Series, Vol. 679,
  Solar Wind Ten, ed. M.~{Velli}, R.~{Bruno}, F.~{Malara}, \& B.~{Bucci},
  660--663

\bibitem[{{Dalla} {et~al.}(2017){Dalla}, {Marsh}, \& {Battarbee}}]{Dal2017a}
{Dalla}, S., {Marsh}, M.~S., \& {Battarbee}, M. 2017, \apj, 834, 167

\bibitem[{{Dalla} {et~al.}(2013){Dalla}, {Marsh}, {Kelly}, \&
  {Laitinen}}]{Dal2013}
{Dalla}, S., {Marsh}, M.~S., {Kelly}, J., \& {Laitinen}, T. 2013, Journal of
  Geophysical Research (Space Physics), 118, 5979

\bibitem[{{Desai} \& {Giacalone}(2016)}]{Des2016}
{Desai}, M. \& {Giacalone}, J. 2016, Living Reviews in Solar Physics, 13, 3

\bibitem[{{Ding} {et~al.}(2022){Ding}, {Li}, {Ebert}, {Dayeh}, {Fe-Due{\~n}as},
  {Desai}, {Xie}, {Gopalswamy}, \& {Bruno}}]{Din2022}
{Ding}, Z., {Li}, G., {Ebert}, R.~W., {et~al.} 2022, Journal of Geophysical
  Research (Space Physics), 127, e30343

\bibitem[{{Domingo} {et~al.}(1995){Domingo}, {Fleck}, \& {Poland}}]{Dom1995}
{Domingo}, V., {Fleck}, B., \& {Poland}, A.~I. 1995, \solphys, 162, 1

\bibitem[{{Dr{\"o}ge} {et~al.}(2010){Dr{\"o}ge}, {Kartavykh}, {Klecker}, \&
  {Kovaltsov}}]{Dro2010}
{Dr{\"o}ge}, W., {Kartavykh}, Y.~Y., {Klecker}, B., \& {Kovaltsov}, G.~A. 2010,
  \apj, 709, 912

\bibitem[{{Forsyth} \& {Gosling}(2001)}]{For2001}
{Forsyth}, R.~J. \& {Gosling}, J.~T. 2001, in The Heliosphere Near Solar
  Minimum. The Ulysses Perspective, ed. A.~{Balogh}, R.~G. {Marsden}, \& E.~J.
  {Smith} (Springer), 107--166

\bibitem[{{Giacalone} \& {Jokipii}(2012)}]{Gia2012}
{Giacalone}, J. \& {Jokipii}, J.~R. 2012, \apjl, 751, L33

\bibitem[{He \& Wan(2017)}]{He&Wan_2017}
He, H.-Q. \& Wan, W. 2017, Monthly Notices of the Royal Astronomical Society,
  464, 85

\bibitem[{{Heber} {et~al.}(1999){Heber}, {Sanderson}, \& {Zhang}}]{Heb1999}
{Heber}, B., {Sanderson}, T.~R., \& {Zhang}, M. 1999, Advances in Space
  Research, 23, 567

\bibitem[{{Hutchinson} {et~al.}(2023){Hutchinson}, {Dalla}, {Laitinen}, \&
  {Waterfall}}]{Hut2023a}
{Hutchinson}, A., {Dalla}, S., {Laitinen}, T., \& {Waterfall}, C.~O.~G. 2023,
  \aap, 670, L24

\bibitem[{{Hyndman} {et~al.}(2024){Hyndman}, {Dalla}, {Laitinen}, {Hutchinson},
  {Cohen}, \& {Wimmer-Schweingruber}}]{Hyn2024}
{Hyndman}, R.~A., {Dalla}, S., {Laitinen}, T., {et~al.} 2024, A\&A, in press,
  DOI: 10.1051/0004

\bibitem[{{Kahler}(2016)}]{Kah2016}
{Kahler}, S.~W. 2016, \apj, 819, 105

\bibitem[{{Kaiser} {et~al.}(2008){Kaiser}, {Kucera}, {Davila}, {St. Cyr},
  {Guhathakurta}, \& {Christian}}]{Kai2008}
{Kaiser}, M.~L., {Kucera}, T.~A., {Davila}, J.~M., {et~al.} 2008, \ssr, 136, 5

\bibitem[{{Kallenrode} {et~al.}(1992){Kallenrode}, {Cliver}, \&
  {Wibberenz}}]{Kal1992b}
{Kallenrode}, M.~B., {Cliver}, E.~W., \& {Wibberenz}, G. 1992, \apj, 391, 370

\bibitem[{{Kallenrode} \& {Wibberenz}(1997)}]{Kal1997}
{Kallenrode}, M.-B. \& {Wibberenz}, G. 1997, \jgr, 102, 22311

\bibitem[{{Kihara} {et~al.}(2020){Kihara}, {Huang}, {Nishimura}, {Nitta},
  {Yashiro}, {Ichimoto}, \& {Asai}}]{Kih2020}
{Kihara}, K., {Huang}, Y., {Nishimura}, N., {et~al.} 2020, \apj, 900, 75

\bibitem[{{Klein} \& {Dalla}(2017)}]{Kle2017}
{Klein}, K.-L. \& {Dalla}, S. 2017, \ssr, 212, 1107

\bibitem[{{Laitinen} {et~al.}(2018){Laitinen}, {Dalla}, {Battarbee}, \&
  {Marsh}}]{Lai2018}
{Laitinen}, T., {Dalla}, S., {Battarbee}, M., \& {Marsh}, M.~S. 2018, in IAU
  Symposium, Vol. 335, Space Weather of the Heliosphere: Processes and
  Forecasts, ed. C.~{Foullon} \& O.~E. {Malandraki}, 298--300

\bibitem[{{Laitinen} {et~al.}(2023){Laitinen}, {Dalla}, {Waterfall}, \&
  {Hutchinson}}]{Lai2023}
{Laitinen}, T., {Dalla}, S., {Waterfall}, C.~O.~G., \& {Hutchinson}, A. 2023,
  \aap, 673, L8

\bibitem[{{Lario} {et~al.}(2013){Lario}, {Aran}, {G{\'o}mez-Herrero},
  {Dresing}, {Heber}, {Ho}, {Decker}, \& {Roelof}}]{Lar2013}
{Lario}, D., {Aran}, A., {G{\'o}mez-Herrero}, R., {et~al.} 2013, \apj, 767, 41

\bibitem[{{Lario} {et~al.}(1998){Lario}, {Sanahuja}, \& {Heras}}]{Lar1998}
{Lario}, D., {Sanahuja}, B., \& {Heras}, A.~M. 1998, \apj, 509, 415

\bibitem[{Marsh {et~al.}(2015)Marsh, Dalla, Dierckxsens, Laitinen, \&
  Crosby}]{Mar2015}
Marsh, M.~S., Dalla, S., Dierckxsens, M., Laitinen, T., \& Crosby, N.~B. 2015,
  Space Weather, 13, 386

\bibitem[{{Mewaldt} {et~al.}(2008){Mewaldt}, {Cohen}, {Cook}, {Cummings},
  {Davis}, {Geier}, {Kecman}, {Klemic}, {Labrador}, {Leske}, {Miyasaka},
  {Nguyen}, {Ogliore}, {Stone}, {Radocinski}, {Wiedenbeck}, {Hawk}, {Shuman},
  {von Rosenvinge}, \& {Wortman}}]{Mew2008a}
{Mewaldt}, R.~A., {Cohen}, C.~M.~S., {Cook}, W.~R., {et~al.} 2008, \ssr, 136,
  285

\bibitem[{{Odstrcil} \& {Pizzo}(1999)}]{Ods1999}
{Odstrcil}, D. \& {Pizzo}, V.~J. 1999, \jgr, 104, 28225

\bibitem[{Onsager {et~al.}(1996)Onsager, Grubb, Kunches, Matheson, Speich,
  Zwickl, \& Sauer}]{Ons1996}
Onsager, T., Grubb, R., Kunches, J., {et~al.} 1996, in GOES-8 and Beyond, ed.
  E.~R. Washwell, Vol. 2812, International Society for Optics and Photonics
  (SPIE), 281 -- 290

\bibitem[{{Owens} {et~al.}(2020){Owens}, {Lang}, {Barnard}, {Riley}, {Ben-Nun},
  {Scott}, {Lockwood}, {Reiss}, {Arge}, \& {Gonzi}}]{Owe2020}
{Owens}, M., {Lang}, M., {Barnard}, L., {et~al.} 2020, \solphys, 295, 43

\bibitem[{Owens {et~al.}(2013)Owens, Challen, Methven, Henley, \&
  Jackson}]{Owe2013}
Owens, M.~J., Challen, R., Methven, J., Henley, E., \& Jackson, D.~R. 2013,
  Space Weather, 11, 225

\bibitem[{{Papaioannou} {et~al.}(2016){Papaioannou}, {Sandberg},
  {Anastasiadis}, {Kouloumvakos}, {Georgoulis}, {Tziotziou}, {Tsiropoula},
  {Jiggens}, \& {Hilgers}}]{Pap2016}
{Papaioannou}, A., {Sandberg}, I., {Anastasiadis}, A., {et~al.} 2016, Journal
  of Space Weather and Space Climate, 6, A42

\bibitem[{{Pesnell} {et~al.}(2012){Pesnell}, {Thompson}, \&
  {Chamberlin}}]{Pes2012}
{Pesnell}, W.~D., {Thompson}, B.~J., \& {Chamberlin}, P.~C. 2012, \solphys,
  275, 3

\bibitem[{{Pomoell} \& {Poedts}(2018)}]{Pom2018}
{Pomoell}, J. \& {Poedts}, S. 2018, Journal of Space Weather and Space Climate,
  8, A35

\bibitem[{Richardson {et~al.}(2014)Richardson, von Rosenvinge, Cane, Christian,
  Cohen, Labrador, Leske, Mewaldt, Wiedenbeck, \& Stone}]{Ric2014}
Richardson, I.~G., von Rosenvinge, T.~T., Cane, H.~V., {et~al.} 2014, Solar
  Physics, 289, 48

\bibitem[{{Rodr{\'\i}guez-Garc{\'\i}a}
  {et~al.}(2023{\natexlab{a}}){Rodr{\'\i}guez-Garc{\'\i}a}, {Balmaceda},
  {G{\'o}mez-Herrero}, {Kouloumvakos}, {Dresing}, {Lario}, {Zouganelis},
  {Fedeli}, {Espinosa Lara}, {Cernuda}, {Ho}, {Wimmer-Schweingruber}, \&
  {Rodr{\'\i}guez-Pacheco}}]{Rod2023b}
{Rodr{\'\i}guez-Garc{\'\i}a}, L., {Balmaceda}, L.~A., {G{\'o}mez-Herrero}, R.,
  {et~al.} 2023{\natexlab{a}}, \aap, 674, A145

\bibitem[{{Rodr{\'\i}guez-Garc{\'\i}a}
  {et~al.}(2023{\natexlab{b}}){Rodr{\'\i}guez-Garc{\'\i}a},
  {G{\'o}mez-Herrero}, {Dresing}, {Lario}, {Zouganelis}, {Balmaceda},
  {Kouloumvakos}, {Fedeli}, {Espinosa Lara}, {Cernuda}, {Ho},
  {Wimmer-Schweingruber}, \& {Rodr{\'\i}guez-Pacheco}}]{Rod2023a}
{Rodr{\'\i}guez-Garc{\'\i}a}, L., {G{\'o}mez-Herrero}, R., {Dresing}, N.,
  {et~al.} 2023{\natexlab{b}}, \aap, 670, A51

\bibitem[{{Sarris} {et~al.}(1984){Sarris}, {Anagnostopoulos}, \&
  {Trochoutsos}}]{Sar1984}
{Sarris}, E.~T., {Anagnostopoulos}, G.~C., \& {Trochoutsos}, P.~C. 1984,
  \solphys, 93, 195

\bibitem[{{Solomon} {et~al.}(2007){Solomon}, {McNutt}, {Gold}, \&
  {Domingue}}]{Sol2007}
{Solomon}, S.~C., {McNutt}, R.~L., {Gold}, R.~E., \& {Domingue}, D.~L. 2007,
  \ssr, 131, 3

\bibitem[{{Strauss} {et~al.}(2017){Strauss}, {Dresing}, \&
  {Engelbrecht}}]{Str2017a}
{Strauss}, R. D.~T., {Dresing}, N., \& {Engelbrecht}, N.~E. 2017, \apj, 837, 43

\bibitem[{{Tylka} {et~al.}(2005){Tylka}, {Cohen}, {Dietrich}, {Lee},
  {Maclennan}, {Mewaldt}, {Ng}, \& {Reames}}]{Tyl2005}
{Tylka}, A.~J., {Cohen}, C.~M.~S., {Dietrich}, W.~F., {et~al.} 2005, \apj, 625,
  474

\bibitem[{{Van Hollebeke} {et~al.}(1975){Van Hollebeke}, {Ma Sung}, \&
  {McDonald}}]{vanHol1975}
{Van Hollebeke}, M.~A.~I., {Ma Sung}, L.~S., \& {McDonald}, F.~B. 1975,
  \solphys, 41, 189

\bibitem[{{von Rosenvinge} {et~al.}(2008){von Rosenvinge}, {Reames}, {Baker},
  {Hawk}, {Nolan}, {Ryan}, {Shuman}, {Wortman}, {Mewaldt}, {Cummings}, {Cook},
  {Labrador}, {Leske}, \& {Wiedenbeck}}]{vRos2008}
{von Rosenvinge}, T.~T., {Reames}, D.~V., {Baker}, R., {et~al.} 2008, \ssr,
  136, 391

\bibitem[{Waterfall {et~al.}(2022)Waterfall, Dalla, Laitinen, Hutchinson, \&
  Marsh}]{Wat2022}
Waterfall, C. O.~G., Dalla, S., Laitinen, T., Hutchinson, A., \& Marsh, M.
  2022, \apj, 934, 82

\bibitem[{{Whitman} {et~al.}(2023){Whitman}, {Egeland}, {Richardson},
  {Allison}, {Quinn}, {Barzilla}, {Kitiashvili}, {Sadykov}, {Bain},
  {Dierckxsens}, {Mays}, {Tadesse}, {Lee}, {Semones}, {Luhmann},
  {N{\'u}{\~n}ez}, {White}, {Kahler}, {Ling}, {Smart}, {Shea}, {Tenishev},
  {Boubrahimi}, {Aydin}, {Martens}, {Angryk}, {Marsh}, {Dalla}, {Crosby},
  {Schwadron}, {Kozarev}, {Gorby}, {Young}, {Laurenza}, {Cliver}, {Alberti},
  {Stumpo}, {Benella}, {Papaioannou}, {Anastasiadis}, {Sandberg}, {Georgoulis},
  {Ji}, {Kempton}, {Pandey}, {Li}, {Hu}, {Zank}, {Lavasa}, {Giannopoulos},
  {Falconer}, {Kadadi}, {Fernandes}, {Dayeh}, {Mu{\~n}oz-Jaramillo},
  {Chatterjee}, {Moreland}, {Sokolov}, {Roussev}, {Taktakishvili},
  {Effenberger}, {Gombosi}, {Huang}, {Zhao}, {Wijsen}, {Aran}, {Poedts},
  {Kouloumvakos}, {Paassilta}, {Vainio}, {Belov}, {Eroshenko}, {Abunina},
  {Abunin}, {Balch}, {Malandraki}, {Karavolos}, {Heber}, {Labrenz}, {K{\"u}hl},
  {Kosovichev}, {Oria}, {Nita}, {Illarionov}, {O'Keefe}, {Jiang}, {Fereira},
  {Ali}, {Paouris}, {Aminalragia-Giamini}, {Jiggens}, {Jin}, {Lee}, {Palmerio},
  {Bruno}, {Kasapis}, {Wang}, {Chen}, {Sanahuja}, {Lario}, {Jacobs}, {Strauss},
  {Steyn}, {van den Berg}, {Swalwell}, {Waterfall}, {Nedal}, {Miteva},
  {Dechev}, {Zucca}, {Engell}, {Maze}, {Farmer}, {Kerber}, {Barnett}, {Loomis},
  {Grey}, {Thompson}, {Linker}, {Caplan}, {Downs}, {T{\"o}r{\"o}k}, {Lionello},
  {Titov}, {Zhang}, \& {Hosseinzadeh}}]{Whi2023}
{Whitman}, K., {Egeland}, R., {Richardson}, I.~G., {et~al.} 2023, Advances in
  Space Research, 72, 5161

\bibitem[{{Wuelser} {et~al.}(2004){Wuelser}, {Lemen}, {Tarbell}, {Wolfson},
  {Cannon}, {Carpenter}, {Duncan}, {Gradwohl}, {Meyer}, {Moore}, {Navarro},
  {Pearson}, {Rossi}, {Springer}, {Howard}, {Moses}, {Newmark},
  {Delaboudiniere}, {Artzner}, {Auchere}, {Bougnet}, {Bouyries}, {Bridou},
  {Clotaire}, {Colas}, {Delmotte}, {Jerome}, {Lamare}, {Mercier}, {Mullot},
  {Ravet}, {Song}, {Bothmer}, \& {Deutsch}}]{Wue2004}
{Wuelser}, J.-P., {Lemen}, J.~R., {Tarbell}, T.~D., {et~al.} 2004, in Society
  of Photo-Optical Instrumentation Engineers (SPIE) Conference Series, Vol.
  5171, Telescopes and Instrumentation for Solar Astrophysics, ed.
  S.~{Fineschi} \& M.~A. {Gummin}, 111--122

\end{thebibliography}

\clearpage

\end{document}